\documentclass[twocolumn,showpacs,preprintnumbers,amsmath,amssymb]{revtex4}
\usepackage{graphicx}% Include figure files
\usepackage{dcolumn}% Align table columns on decimal point
\usepackage{bm}% bold math
\usepackage{epsfig}
\begin{document}

\title{Liquid Polymorphism and Density Anomaly in a Lattice Gas Model}

\author{Vera B. Henriques}
\affiliation{Instituto de Física, Universidade de S\~ao Paulo,
Caixa Postal 66318, 05315970, São Paulo, SP, Brazil}
\email{vhenriques@if.usp.br}
\author{Marcia C. Barbosa}
\affiliation{Instituto de F\'{\i}sica, UFRGS, Caixa Postal 15051, 91501-970, Porto Alegre, RS, Brazil}
\email{barbosa@if.ufrgs.br}
\homepage{http://www.if.ufrgs.br/~barbosa}

\date{\today}
\begin{abstract}

We present a simple model for an associating liquid in which polymorphism and
density anomaly are connected. Our model combines a two dimensional lattice
gas with particles interacting through a soft core potential and orientational
degrees of freedom represented through thermal \char`\"{}ice variables\char`\"{}.
The competition between the directional attractive forces and the soft core
potential leads to a phase diagram in which two liquid phases and a density
anomaly are present. The coexistence line between the low density liquid and
the high density liquid has a positive slope contradicting the surmise that
the presence of a density anomaly implies that the high density liquid is more
entropic than the low density liquid.

\end{abstract}

\pacs{64.70.Ja, 05.70.Ce, 05.10.Ln }
\maketitle

Water is one of the most mysterious materials in nature. It exhibits a number
of thermodynamic and dynamic anomalous properties \cite{Fr85}-\cite{Gr04},
such as the maximum as a function of temperature both in density and in isothermal
compressibility in the liquid phase. It has been proposed some time ago that
these anomalies might be associated with a critical point at the terminus of
a liquid-liquid line, in the unstable supercooled liquid region\cite{Po92},
at high pressures, following the suggestion, based on varied experimental data
\cite{angell}, of a thermodynamic singularity in supercooled water, around
\( 228K \) and at atmospheric pressure. Inspite of the limit of \( 235K \)
below which water cannot be found in the liquid phase without crystallization,
two amorphous phases were observed at much lower temperatures \cite{Mi84}.
There is evidence, although yet under test, that these two amorphous phases
are related to fluid water \cite{Sm99}\cite{Mi02}. 

Notwithstanding of its confirmation for metastable water, interest in liquid
polymorphism arose, and the coexistence of two liquid phases was uncovered as
a possibility for a few other both associating and non-associating liquids.
Notable examples include liquid metals \cite{Cu81}, silica \cite{lacks}, phosphorus
\cite{poli2}\cite{key-1} and graphite \cite{poli1}. The relation between
liquid polymorphism and density anomaly has been a subject of debate in recent
theoretical literature \cite{Fr01}.

From a microscopic point of view, water anomalies have been interpreted qualitatively,
since\cite{bernal}, in terms of the the presence of an extensive hydrogen bond
network which persists in the fluid phase \cite{errington}. The hydrogen bond
net deformation under temperature was represented in a number of minimal models
for water. The main strategy has been to associate the hydrogen bond disorder
with bond \cite{sastry}\cite{giancarlo} or site \cite{sastry_93}\cite{debenedetti}
Potts states. In the former case coexistence between two liquid phases may follow
from the presence of an order-disorder transition and a density anomaly is introduced
\emph{ad hoc} by the addition to the free energy of a volume term proportional
to a Potts order parameter. In the second case, it may arise from the competition
between occupational and Potts variables introduced through a depedency of bond
strength on local density states. 

We propose a description also based on occupational and orientational degrees
of freedom. However, for the orientational part we employ a modification of
the thermal version\cite{hu83}\cite{Na91} of the ice model\cite{lieb}, so successful in
the description of ice entropy. Competition between the filling up of the lattice
and the formation of an open four-bonded orientational structure is naturally
introduced in terms of the ice bonding variables and no \emph{ad hoc} introduction
of density or bond strength variations is needed. We thus consider a lattice
gas on a triangular lattice with sites which may be full or empty. Besides the
occupational variables, \( \sigma _{i} \), associated to each particle \( i \)
there are six other variables, \( \tau _{i}^{ij} \), pointing to neighboring
sites \( j \): four are the usual ice bonding arms, two donor, with \( \tau _{i}^{ij}=1 \),
and two acceptor, with \( \tau _{i}^{ij}=-1 \), while two additional opposite
arms are taken as inert (non-bonding), \( \tau _{i}^{ij}=0 \), as illustrated
in Fig. 1. Therefore each occupied site is allowed to be in one of eighteen
possible states. 
%%%%%%%%%%%%%%%%%%%%%%%%%%%%%%%%%%%%%%%%%%%%%%%%%%%%%%%%%%%%%%%%
\begin{figure}
{par\centering \resizebox*{6cm}{4cm}{\includegraphics{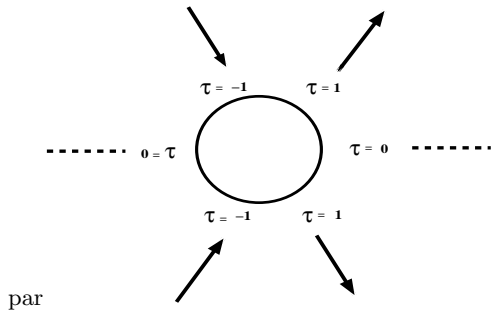}} \par}
\caption{The model orientational state: four bonding (donor and receptor) and two non-bonding arms}
\label{fig1}
\end{figure}
%%%%%%%%%%%%%%%%%%%%%%%%%%%%%%%%%%%%%%%%%%%%%%%%%%%%%%%%%%%%%%%%%

Two kinds of interactions are considered: isotropic ``van der Waals'' and
orientational hydrogen bonding. An energy \( -v \) is attributed to each pair
of occupied neighboring sites that form a hydrogen bond, while non-bonding pairs
have an energy, \( -v+2u \) (for \( u>0 \)). The overall model energy is given
by 
%%%%%%%%%%%%%%%%%%%%%%%%%%%%%%%%%%%%%%%%%%%%%%%%%%%%%%%%%%%%%%%%%%%%%55
\begin{equation}
\label{E}
E=\sum _{(i,j)}\sigma _{i}\sigma _{j}\{-v+u[2+\tau ^{ij}_{i}\tau ^{ji}_{j}(1-\tau ^{ij}_{j}\tau ^{ji}_{j})]\}
\end{equation}
%%%%%%%%%%%%%%%%%%%%%%%%%%%%%%%%%%%%%%%%%%%%%%%%%%%%%%%%%%%%%%%%%%%%%%%%55
 where \( \sigma _{i}=0,1 \) are occupation variables and \( \tau _{i}^{ij}=0,\pm 1 \)
represent the arm states described above.

Inspection of the model properties allows the prediction of two ordered states,
as shown in Fig. 2. One of them has lower density ,\( \rho =0.75 \), and energy
density given by \( e=E/N=-3v/2 \). The other state has higher density ,\( \rho =1 \),
and energy density \( e=-3v+2u \). The addition of an external chemical potential
\( \mu  \) may favor one or the other of the two ordered states. At zero temperature,
the low density liquid (LDL) coexists with the high density liquid (HDL) at
chemical potential \( \mu /v=-6+8u/v \), obtained by equating the grand potential
density (or pressure) associated with each one of these phases. Similarly the coexistence 
pressure at the zero temperature  is given by \( p/v=-3+6u/v \).
Besides these two liquid states, a gas phase is also found and it coexists with
the low density liquid at chemical potential \( \mu /v=-2 \) and pressure \( p=0 \).
The condition for the presence of the two liquid phases is therefore \( u/v>0.5 \).

%%%%%%%%%%%%%%%%%%%%%%%%%%%%%%%%%%%%%%%%%%%%%%%%%%%%%%%%%%%%%%%%%%%%%%%%%
\begin{figure}
{\par\centering \resizebox*{6cm}{4cm}{\includegraphics{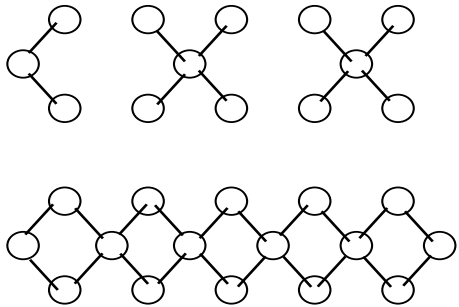}} \par}

\caption{Low density liquid, \protect\( LDL\protect \) with density 3/4 (top), and high density liquid, \protect\( HDL\protect \) witn density 1 (botton) on the triangular lattice.}
\label{fig2}
\end{figure}

%%%%%%%%%%%%%%%%%%%%%%%%%%%%%%%%%%%%%%%%%%%%%%%%%%%%%%%%%%%%%%%%%%%%%%%%%

The model properties for finite temperatures were obtained through Monte Carlo
simulations in the grand-canonical ensemble using the Metropolis algorithm.
Particle insertion and exclusion were tested with transition probabilities given
by 
%%%%%%%%%%%%%%%%%%%%%%%%%%%%%%%%%%%%%%%%%%%%%%%%%%%%%%%%%%%%%%%%%%%%%%%%
\begin{eqnarray}
\label{w}
w(insertion)&=&exp(-\Delta \phi )\;\; and  \nonumber \\
w(exclusion)&=&1\; \; \; if\quad \Delta \phi >0 \nonumber \\
&&or \nonumber \\
w(insertion)&=&1\; \; and  \nonumber \\
w(exclusion)&=&exp(+\Delta \phi )\; \; \; if\quad \Delta \phi <0
\end{eqnarray}
%%%%%%%%%%%%%%%%%%%%%%%%%%%%%%%%%%%%%%%%%%%%%%%%%%%%%%%%%%%%%%%%%%%%%%%%
 with \( \Delta \phi \equiv \exp \{\beta (e_{particle}-\mu )-\ln (18)\} \) where $e_{particle}$ is the energy of the particle included.
Since the empty and full sites are visited randomly, the factor 18 is required
in order to guarantee detailed balance.

Simulational data were generated both from fixed temperature and fixed chemical
potential simulations. Some test runs were done for L=4, 10 and 20. A detailed
study of the model properties and the full phase diagrams was undertaken for
an L=10 lattice, at fixed \( u/v=1 \)\cite{future}. Runs were of the order
of \( 10^{6} \) Monte Carlo steps.

The three phases obtained at zero temperature are present for low temperatures,
as can be seen in the isotherms of Figure 3. The model exhibits two first order
phase transition lines, gas-LDL and LDL-HDL, respectively.

In order to obtain the complete phase diagram, including the two critical points,
and to check for density anomalies, pressure was computed by numerical integration
of the Gibbs Duhem equation, \( SdT-VdP+Nd\mu =0 \), at fixed temperature.
Integration was carried out from effective zero density, at which pressure is
zero, to obtain \( P(\rho ,T) \) isotherms.

%%%%%%%%%%%%%%%%%%%%%%%%%%%%%%%%%%%%%%%%%%%%%%%%%%%%%%%%%%%%%%%%%%%%%%%%%
\begin{figure}
\begin{minipage}[b]{.8\linewidth}
\centering \epsfig{figure=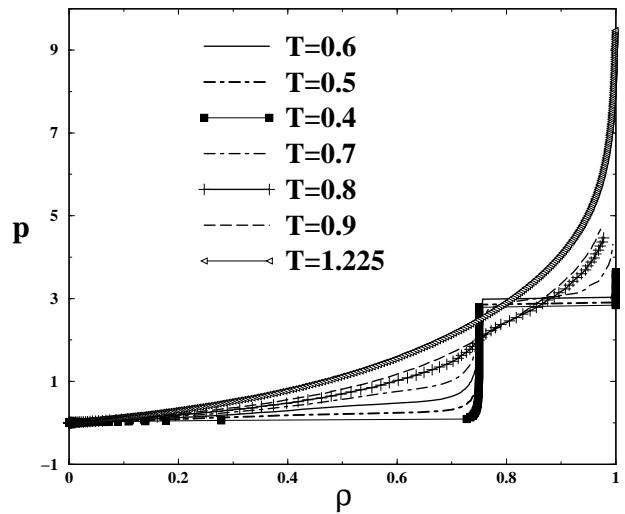,width=\linewidth,angle=-90}
\caption{Pressure vs. density isotherms for different temperatures.}
\label{fig3}
\end{minipage} \hspace{.8cm}
\end{figure}

%%%%%%%%%%%%%%%%%%%%%%%%%%%%%%%%%%%%%%%%%%%%%%%%%%%%%%%%%%%%%%%%%%%%%%%%%
The pressure isotherms show that an inversion of the behavior
of density as a function of temperature takes place at intermediate pressures,
in the LDL phase. At smaller pressures ,\( p\sim 1 \), density decreases with
temperature, whereas at higher pressures ,\( p\sim 3 \), density increases
with temperature. This yields a density anomaly in the higher range of pressures,
which we illustrate in Fig 4.
%%%%%%%%%%%%%%%%%%%%%%%%%%%%%%%%%%%%%%%%%%%%%%%%%%%%%%%%%%%%%%%%%%%%%%%%%
\begin{figure}
\begin{minipage}[b]{.8\linewidth}
\centering \epsfig{figure=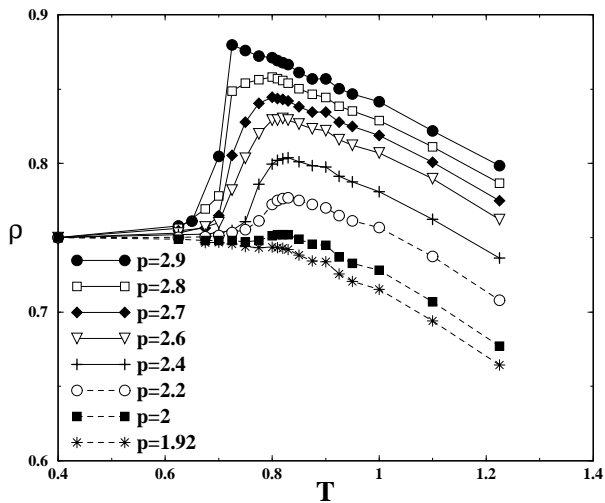,width=\linewidth,angle=-90}

\caption{Density Anomaly for different pressures}
\label{fig4}
\end{minipage} \hspace{.8cm}
\end{figure}

%%%%%%%%%%%%%%%%%%%%%%%%%%%%%%%%%%%%%%%%%%%%%%%%%%%%%%%%%%%%%%%%%%%%%%%%%

%%%%%%%%%%%%%%%%%%%%%%%%%%%%%%%%%%%%%%%%%%%%%%%%%%%%%%%%%%%%%%%%%%%%%%%%%
\begin{figure}
\begin{minipage}[b]{.8\linewidth}
\centering \epsfig{figure=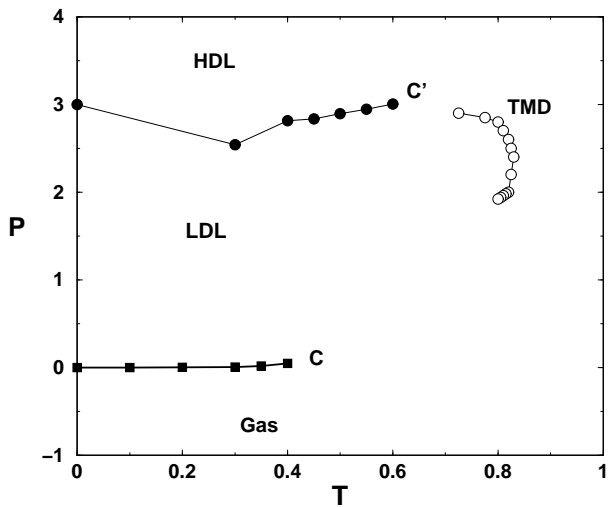,width=\linewidth,angle=-90}

\caption{Phase Diagram. The open circles represent the temperature of maximum density, $TMD$, line, the filled circles represent the LDL-HDL  coexistence that ends at the critical point $C'$ and the filled squares represent the gas-LDL coexistence line ending at the critical point C.}
\label{fig5}
\end{minipage} \hspace{.8cm}
\end{figure}

%%%%%%%%%%%%%%%%%%%%%%%%%%%%%%%%%%%%%%%%%%%%%%%%%%%%%%%%%%%%%%%%%%%%%%%%%

Finally, from a large set of temperatures, we build up the pressure versus temperature
coexistence curves shown in Fig. 5. The line of maximum densities is also shown.
The liquid-liquid coexistence line has a positive inclination, except at very
low temperatures (the zero temperature points do not come from simulations,
but from equating the enthalpy densities). From the Clapeyron condition, \( \frac{dp}{dT}]_{coex}=\frac{\Delta s}{\Delta v} \),
the positive slope implies that within our model the HDL phase has lower entropy
than the LDL phase.

Inside each phase, the density anomaly can be related to the behavior of entropy
as a function of pressure. From thermodynamics, a negative thermal expansion
coefficient \( \alpha \equiv (\frac{\partial v}{\partial T})_{p} \)implies
a positive gradient of entropy with pressure, since \( (\frac{\partial v}{\partial T})_{p}=-(\frac{\partial s}{\partial p})_{T} \).
This property has been thought \cite{Fr01} to imply that the presence of a
density anomaly would lead to a high entropy high density phase, and therefore
to a negative slope of the coexistence line, as is true for the ice fusion line.
The present model proves that this assumption is misfounded and that this is
not a general behavior.

What we have here is the following: on the low density side, the thermal expansion
coefficient is negative, whereas on the high density phase it is positive, as
can be gathered from the pressure-density isotherms. The positive slope of the
coexistence line implies, by Clausius-Clapeyron, that the high density phase
is the lower entropy phase. Thus, at constant temperature, entropy increases
with pressure up to the coexistence line, drops discontinuously across this
line, and then decreases with pressure, as in any normal liquid. Therefore the
sign of entropy variation across the coexistence line may be either positive,
as in this model, or negative, as in the fusion of ice, following, in \emph{both
cases}, the high pressure \( \alpha  \) sign.

The model proposed is a truly statistical model which includes orientational
and occupational variables, and guarantees the local distribution of hydrogens
on molecular bonds, without the need of increasing the volume artificially or
introducing artificial orientational variables. Inspite of the absence of an
orientational order-disorder transition \cite{Na91}, the model presents liquid-liquid coexistence, with positive inclination in the pressure-temperature plane, accompanied
by a line of maximum density, on the low density side, a feature expected for
water. Besides, this study points out to the fact that the presence of a density
anomaly, with \( \alpha <0 \), on the low temperature side, and as a consequence,
\( (\frac{\partial S}{\partial p})_{T}>0 \), \emph{does not} imply a negative
slope of the liquid-liquid line, contrasting with the results for most studies
of metastable liquid-liquid coexistence in models for water, which suggest a
transition line with negative gradient \cite{Sc00}.

The presence of both a density anomaly and two liquid phases in our model begs
the question of which features of this potential are responsible for such behaviour.
Averaged over orientational degrees of freedom, our model can be seen as some
kind of shoulder potential, with the liquid-liquid coexistence line being present
only for a repulsive, soft-core potential. The same was indeed observed for
continuous step pair potentials \cite{Fr01}\cite{Fr02}, for which, however,
the density anomaly is absent. On the other hand, a density anomaly seems to
be associated with smooth soft core potentials \cite{Ja01}\cite{Wi02}, which
would be hidden, in our model, in the orientational degrees of freedom.

In summary, we have found that a lattice gas with orientational ice-like degrees
of freedom can generate a density anomaly and a liquid-liquid phase boundary
with positive slope.

\vspace*{1.25cm}

\noindent \textbf{\large Acknowledgments}{\large \par}

\vspace*{0.5cm} This work was supported by the brazilian science agencies CNPq,
FINEP, Fapesp and Fapergs.


\begin{thebibliography}{10}
\bibitem{future}Results for the model have been presented for a single set of model parameters.
The temperature of the second critical point decreases and becomes negative
as \( v/u\rightarrow 0.5 \).
\end{thebibliography}
\end{document}